# Seamless Data Services for Real Time Communication in a Heterogeneous Networks using Network Tracking and Management

**Adiline Macriga. T[1]** **Dr. P. Anandha Kumar[2]**

1. Research Scholar, Department of Information & Communication, MIT Campus, Anna University Chennai, Chennai – 600025. email: adiline_shanker@rediffmail.com, adiline.it@ssec.edu.in
2. Asst. Professor, Department of Information Technology, MIT Campus, Anna University Chennai, Chennai – 600025. email: anandh@annauniv.edu

**Abstract**

*Heterogeneous Networks is the integration of all existing networks under a single environment with an understanding between the functional operations and also includes the ability to make use of multiple broadband transport technologies and to support generalized mobility. It is a challenging feature for Heterogeneous networks to integrate several IP-based access technologies in a seamless way. The focus of this paper is on the requirements of a mobility management scheme for multimedia real-time communication services - Mobile Video Conferencing. Nowadays, the range of available wireless access network technologies includes cellular or wide-area wireless systems, such as cellular networks (GSM/GPRS/UMTS) or Wi-Max, local area Network or personal area wireless systems, comprising for example, WLAN (802.11 a/b/g) and Bluetooth. As the mobile video conferencing is considered, the more advanced mobile terminals are capable of having more than one interface active at the same time. In addition, the heterogeneity of access technologies and also the seamless flow of information will increase in the future, making the seamless integration of the access network a key challenge for mobility management in a heterogeneous network environment. Services must be provided to the user regardless of the particular access technology and also the type of service provider or the network used.*

## I. INTRODUCTION

Today's communication technology becomes outdated for tomorrows requirement. The growth of the communication industry is tremendous and unimaginable. There are different modes like "wired, wireless, adhoc, mobile etc., supporting the growth of the communication industry but with all certain limits. Now, it is time to emerge into the world of mobility where the wireless communication plays a vital role, where it is necessary to satisfy the requirements of the modern world. A world without a mobile phone is unimaginable. It has taken people to a different world. Now it is the time for providing services in an uninterrupted way. In medical industry a millisecond delay in transfer of information may lead to a loss of life. So the technology has to be updated day by day to meet the needs of the various industries. As the communication industry is considered it is one of the challenging one for the researchers. Considering the infrastructure of the existing communication industry a huge amount has been deployed by different service providers to satisfy their customer needs. It is now a challenge to provide the flow of information seamlessly without re-modifying the existing infrastructure. The most challenging one is to provide mobility management for real time communication services. This paper mainly concentrates on framing a path in advance between the source and the destination based on the nature of the data being communicated in this case mobile video conferencing. This paper discusses the challenge involved in the mobility and handoff technique. In heterogeneous wireless networks, traditionally handoff is mainly classified as : horizontal handoff and vertical handoff. A horizontal handoff is made between different access points within the same link-layer technology such as when transferring a connection from one Base Station to another or from one Access Point to another. A vertical handoff is a handoff between access networks with different link-layer technologies, which will involve the transfer of a connection between a Base Station and an Access Point. Seamless and efficient Vertical Handoff between different access technologies is an essential and challenging problem in the development toward the next-generation wireless networks [1][12]. Internally as the handoff process is considered it can be further carried out using the following main steps: system discovery, handoff decision, and handoff execution [24]. During the system discovery phase, mobile terminals equipped with multiple interfaces have to determine which networks can be used and the services available in each network. During the handoff decision phase, the mobile device determines which network it should connect to. During the handoff execution phase, connections need to be rerouted from the existing network to the new network in a seamless manner. There are three strategies for handoff decision mechanisms: mobile-controlled handoff, network-controlled handoff, and mobile-assisted handoff [14]. Mobile Controlled Handoff is used in IEEE 802.11 WLAN networks, where a Mobile Host continuously monitors the signal of an Access Point and initiates the handoff procedure. Network Controlled Handoff is used in cellular voice networks where the decision mechanism of handoff control is located in a network entity. Mobile Assisted Handoff has been widely adopted in the current WWANs such as GPRS, where the mobile host measures the signal of surrounding base stations and the network then employs this information and decides whether or not to trigger handoff [3][13]. The handoff algorithms[3] considered are based on the threshold comparison of one or more metrics and dynamic programming/artificial intelligent techniques applied to improve the accuracy of the handoff procedure. The most common metrics are received signal strength, carrier-to-interference ratio, signal-to-interference ratio, and bit error rate [2]. In heterogeneous wireless networks, even though the functionalities of access networks are different, all the networks use a specific signal beacon or reference channel with a constant transmit power to enable received signal strength measurements. Thus, it is very natural and reasonable for vertical handoff algorithms to use received signal strength as the basic criterion for handoff decisions [14] [16].

In order to avoid the ping-pong effect [4], additional parameters such as hysteresis and dwelling timer can be used solely or jointly in the handoff decision process. In addition to the absolute received signal strength threshold, a relative received signal strength hysteresis between the new base station and the old base station is added as the handoff trigger condition to decrease unnecessary handoffs. With this implementation it will be possible to provide an uninterrupted flow of multimedia communication. The ultimate goal of this paper is to provide the services based on the saying "Any time Anything Anywhere and Everywhere Anystate and Everystate". We follow this approach by proposing a location tracking based solution that supports hybrid handovers without disruption of real time multimedia communication services. The solution provided is mobility management using Location based tracking and Network management.

The rest of the paper is organized as follows: Section II reviews related work on location estimation and vertical handover. Section III provides the solution for mobility management. Section IV presents the proposed method. Section V deals with the performance evaluation. Section VI presents the results and related discussion. Section VII discusses directions for future work and concludes this paper.

## II. SURVEY OF EXISTING TECHNOLOGIES & METHODS

The session below provides the survey on the available technologies that supports the communication at the various stages and aspects. Daniel et al. [5] proposes a handoff scheme based on RSS with the consideration of thresholds and hysteresis for mobile nodes to obtain better performance. However, in heterogeneous wireless networks, RSS from different networks can vary significantly due to different techniques used in the physical layers and cannot be easily compared with each other. Thus, the methods in [4] and [5] cannot be applied to VHO directly. Anthony et al. [19] use the dwelling timer as a handoff initiation criterion to increase the WLAN utilization. It was shown in [21] that the optimal value for the dwelling timer varies along with the used data rate or, to be more precise, with the effective throughput ratio. In [8], Olama et al. extend the simulation framework in [14] by introducing a scenario for multiple radio network environments. Their main results show that the handoff delay caused by frequent handoff has a much bigger degrading effect for the throughput in the transition region. In addition, the benefit that can be achieved with the optimal value of the dwelling timer as in [7] may not be enough to compensate for the effect of handoff delay. In [27], claudio et al. propose an automatic Interface selection approach by performing the VHO if a specific number of continuous received beacons from the WLAN exceed or fall below a predefined threshold.

Additionally, in the real-time service, the number of continuous beacon signals should be lower than that of the non-real-time service in order to reduce the handoff delay [26][30]. More parameters may be employed to make more intelligent decisions. Li et al. [10] propose a bandwidth aware VHO technique which considers the residual bandwidth of a WLAN in addition to RSS as the criterion for handoff decisions. However, it relies on the QBSS load defined in the IEEE 802.11e Standard to estimate the residual bandwidth in the WLAN. In [29], Weishen et al. propose a method for defining the handoff cost as a function of the available bandwidth and monetary cost. In[16], actual RSS and bandwidth were chosen as two important parameters for the Waveform design. Hossain et al. [15] propose a game theoretic frame work for radio resource management perform VHO in heterogeneous wireless networks. One main difficulty of the cost approach is its dependence on some parameters that are difficult to estimate, especially in large cellular networks. Mohanty and Akyildiz [14] developed a cross-layer (Layer 2 + 3) handoff management protocol CHMP, which calculates a dynamic value of the RSS threshold for handoff initiation by estimating MH's speed and predicting the handoff signaling delay of possible handoffs.

To sum up, the application scenario of current Vertical Handoff algorithms is relatively simple. For example, most Vertical Handoff algorithms only consider the pure Vertical Handoff scenario, where the algorithm only needs to decide when to use a 3G network and when to use a WLAN [1], [10], [17], [18], [21], [25]. In fact, at any moment, there may be many available networks (homogeneous or heterogeneous), and the Handoff algorithm has to select the optimal network for Horizontal Handoff or Vertical Handoff from all the available candidates. For example, if the current access network of Mobile Host is a WLAN, the Mobile Assisted Handoff may sense many other WLANs and a 3G network at a particular moment, and it has to decide whether to trigger Horizontal Handoff or Vertical Handoff. If the Horizontal Handoff trigger is selected, Mobile assisted handoff then needs to decide which WLAN is the optimal one [20] [22]. Consequently, an analytical framework to evaluate VHO algorithms is needed to provide guidelines for optimization of handoff in heterogeneous wireless networks. It is also necessary to build reasonable and typical simulation models to evaluate the performance of VHO algorithms. The proposed work will provide an optimal solution to choose the network based on the type of service that is being carried out. As the work mainly concentrates on the application oriented approach the type of network and handover selection is also based on requirement It also concentrates on the identification of the source and destination and by locating them the path is chosen.

## III. SOLUTIONS FOR MOBILITY MANAGEMENT

### A. HYBRID HANDOFF

Seamless and efficient information flow between different access technologies is an essential and challenging problem in the development toward the next-generation wireless networks. Also a single hand off cannot be suggested. So the available handoff for different technologies is grouped under the single head hybrid handoff. For the seamless flow of information the main technique is handoff.In general, the handoff process can be divided into three main steps: system discovery, handoff decision, and handoff execution. Also the During the system discovery phase, mobile terminals equipped with multiple interfaces have to determine which networks can be used and the services available in each network. During the handoff decision phase, the mobile device determines which network it should connect to. During the handoff execution phase, connections need to be rerouted from the existing network to the new network in a seamless manner. During the Vertical Handoff procedure, the handoff decision is the most important step that affects Mobile Host's communication [9][11]. An incorrect handoff decision may degrade the QoS of traffic and even break off current communication. Handoff algorithms in heterogeneous wireless networks should support both Horizontal Handoff and Vertical Handoff and can trigger Horizontal handoff or Vertical handoff based on the network condition. What should be noted is that, because of the uncertainty of the network distribution and the randomness of MH's mobility, it is impossible to forecast the type of the next handoff in advance. For this purpose only in the proposed work the nature of the network between the source and the destination is studied in advance and the triggering based on the availability and services provided by the available network. Thus, handoff algorithms in heterogeneous wireless networks must make the appropriate handoff decision based on the network metrics in a related short time scale.

There are three strategies for handoff decision mechanisms: mobile-controlled handoff, network-controlled handoff, and mobile-assisted handoff [14]. Mobile-controlled handoff is used in IEEE 802.11 WLAN networks, where the Mobile Host continuously monitors the signal of an AP and initiates the handoff procedure. Network-controlled handoff is used in cellular voice networks where the decision mechanism of handoff control is located in a network entity. Mobile-controlled handoff has been widely adopted in the current WWANs such as GPRS, where the Mobile handoff measures the signal of surrounding Base Stations and the network then employs this information and decides whether lor not to trigger handoff. During Vertical handoff, only Mobile Hosts have the knowledge about what kind of interfaces they are equipped with. Even if the network has this knowledge, there may be no way to control another network that the Mobile Host is about to handle the handoff . This is achieved by having a proper QoS understanding between the networks. Eg. The data considered for transmission is huge and the available network has a bandwidth that can handle the data but dute to the flow density in the network it is not able to handle that. The proposed work will take the nature of the data and if it is an emergency information which requires immediate attention the corresponding network will be requested to rearrange the traffic in the network by sharing the resources of the surrounding network and to provide the network for the current transmission.

### B. MOBILE INTERNET PROTOCOL

Mobile Internet Protocol is a mobility solution working at the network layer. IPv4 assumes that every node has its own IP address that should remain unchanged during a communication session. Mobile IP introduces the concepts of home address, the permanent address of the Mobile Host and of care-of-address. The latter is a temporary address assigned to the Mobile Host as soon as it moves from its home network to a foreign one. A specific router in the home network in the home agent is informed as soon as the node acquires the Care-of-Address in the foreign network from a so-called foreign agent. The home agent acts as an anchor point, relaying the packets addressed to the home address towards the actual location of the Mobile Host, at the care-of-address. Using mobile IP for real-time communications has some drawbacks. A well-known problem is triangular routing, that is, the fact that the packets sent to the Mobile Host are captured by the home agent and tunneled, whereas the Mobile Host can send packets directly to the Corresponding Host. This asymmetric routing adds delay to the traffic towards the Mobile Host, and delay is an important issue in voice over IP (VoIP). The fact that the packets are tunneled also means that an overhead of typically 20 bytes, due to the IP encapsulation, will be added to each packet. Still another drawback of using mobile IP is that each Mobile Host requires a permanent home IP address, which can be a problem because of the limited number of IP addresses in IPv4. A number of works have built upon MIP to overcome its drawbacks. A notable one is cellular IP [4], which improves MIP, providing fast handoff control and paging functionality comparable to those of cellular networks. Being a network level solution, cellular IP requires support from the access networks, and it is suitable for micro-mobility, namely, mobility within the environment of a single provider. The major work of this paper concentrates on seamless streaming in heterogeneous networks; one of the major challenges is the bit rate adaptation when the gap between two different networks is large. In the heterogeneous networks, an available channel bandwidth usually fluctuates in a wide range from bit rate below 64kbps to above 1mbps according to the type of network. The following technique will overcome the above mentioned drawbacks and will provide a comfortable solution for the seamless flow of information based on the nature of information.

## IV. PROPOSED SERVICE METHOD

Given the requirements for seamless mobility listed above, the roles to be supported by a mobility management function exploiting the terminal capability to access several radio access networks are the following:

i. Selection of the access network at application launch. This role is ensured by mobility management subfunctions here referred to as *service-to-radio mapping control*.

ii. Triggering of the handover during a session. The mobility management function aims at always providing the best access network to the terminal.

iii. A terminal-centric selection without network assistance or recommendation

iv. A network-controlled selection within network entities, based on both terminal and access network measurements, enforcing decisions on the terminal

v. Network-assisted selection on the terminal side, the network providing operator policies, and access/core load information (joint terminal/network decisions). When only one access remains available, network-assisted selection is applied; when access selection is triggered by network load considerations, network control may be used for load balancing.

vi. Finally, for access network selection, the mobility management function must retrieve the status of resource usage in each access network. This information is provided by an “access network resource management” function, which computes a technology-independent abstracted view of access resource availability.

***Functional Entities Involved in Mobility Management*** — The mobility management functions are supported by functional entities described below that are responsible for selecting the best access network for each terminal. They may be triggered at application launch or during the time of connection establishment.

- Generate a geographical map between the source and the destination based on the mode of transport
- Tabulate the list of service providers with their frequency and coverage
- Create an acyclic graph by pointing out the service providers and the geographical scenarios
- Form a topological graph from the acyclic graph by removing the obstacles and considering the signal strength
- Get the traffic status of each and every link from source to destination
- Now create a path from source to destination based on the network traffic and also by choosing the shortest path with minimum handovers

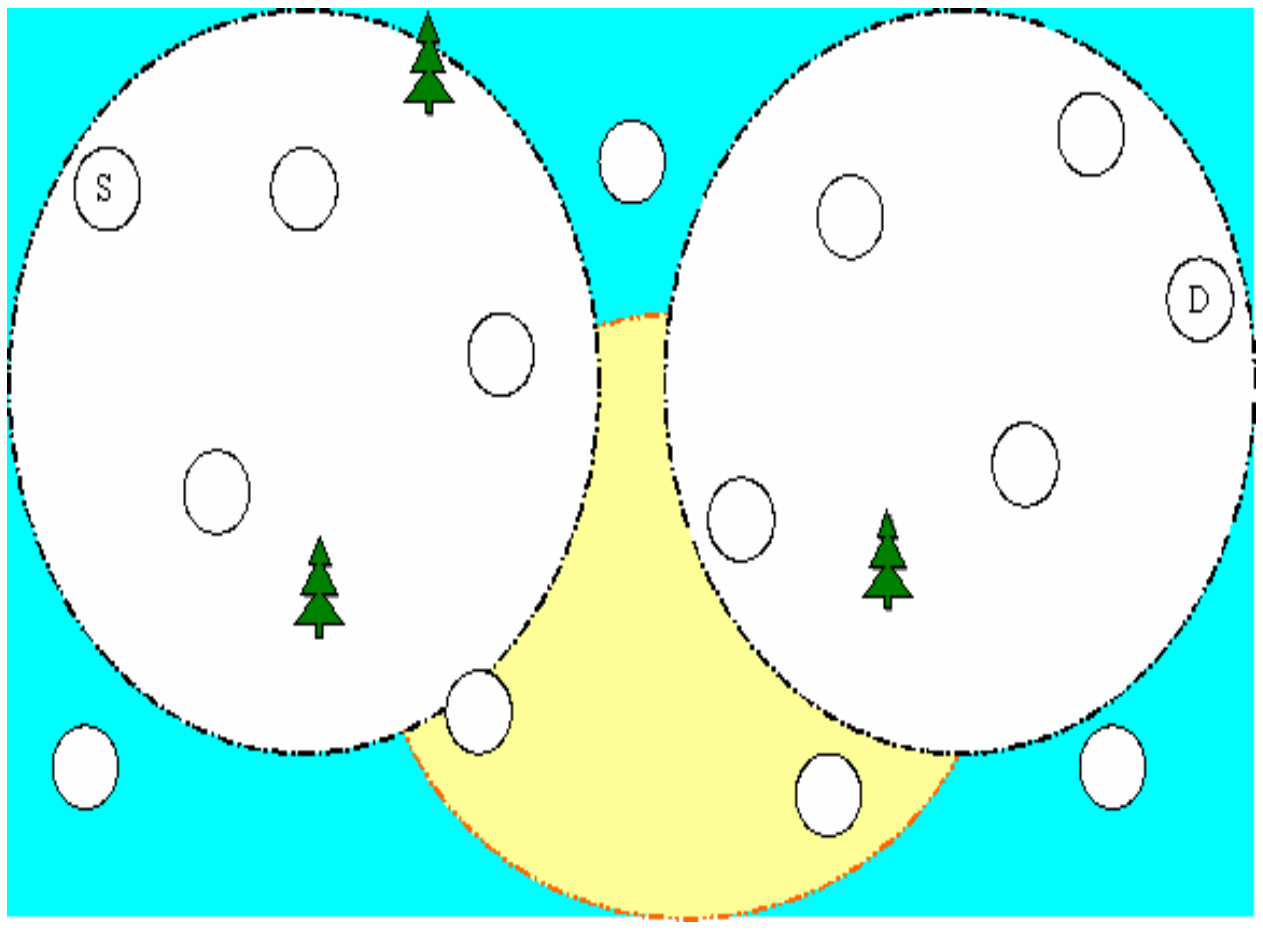

*Fig. 2 Graphical Representation between source (S) and Destination (D)*

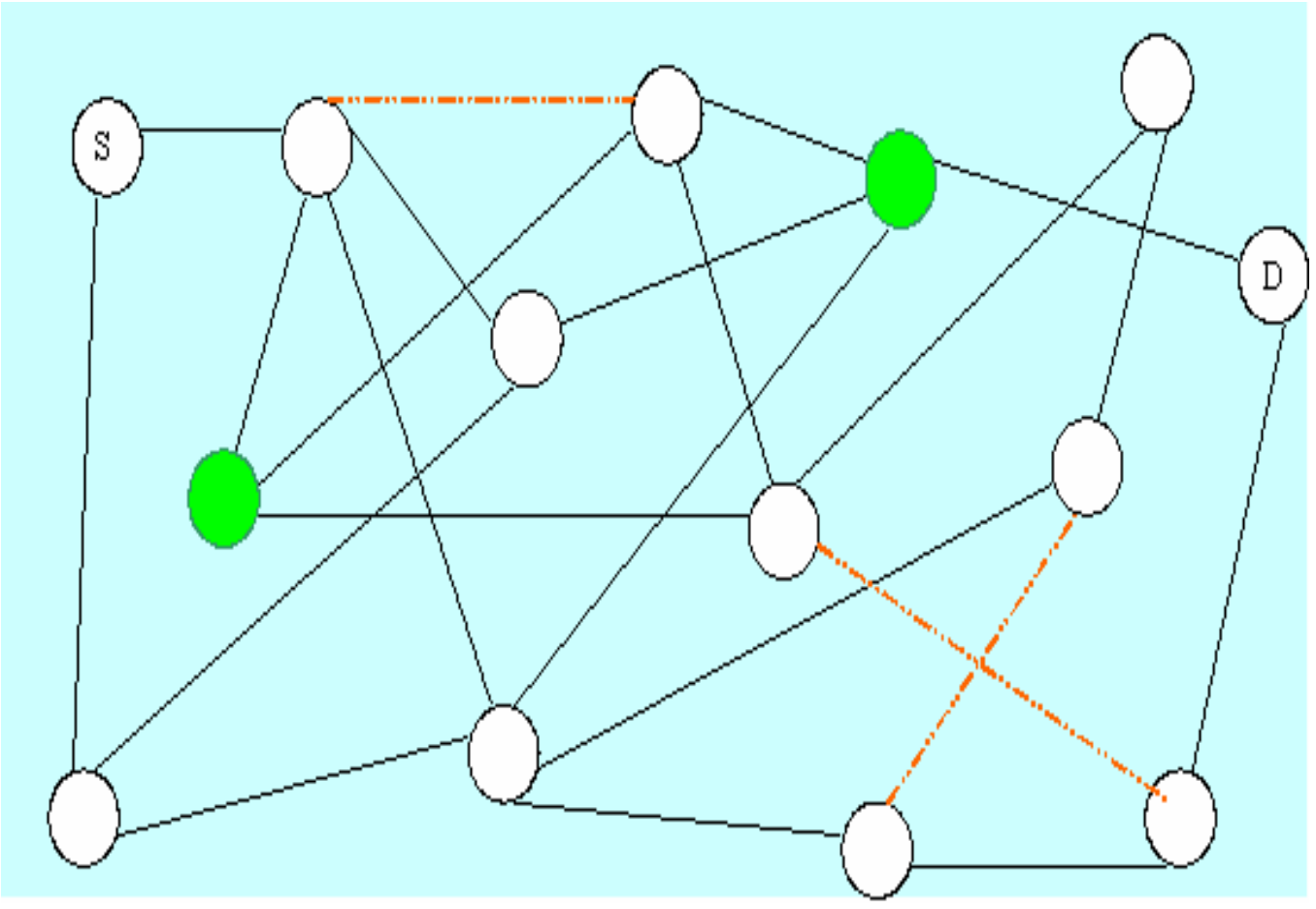

*Fig.3Topological Representation between source (S) and Destination (D)*

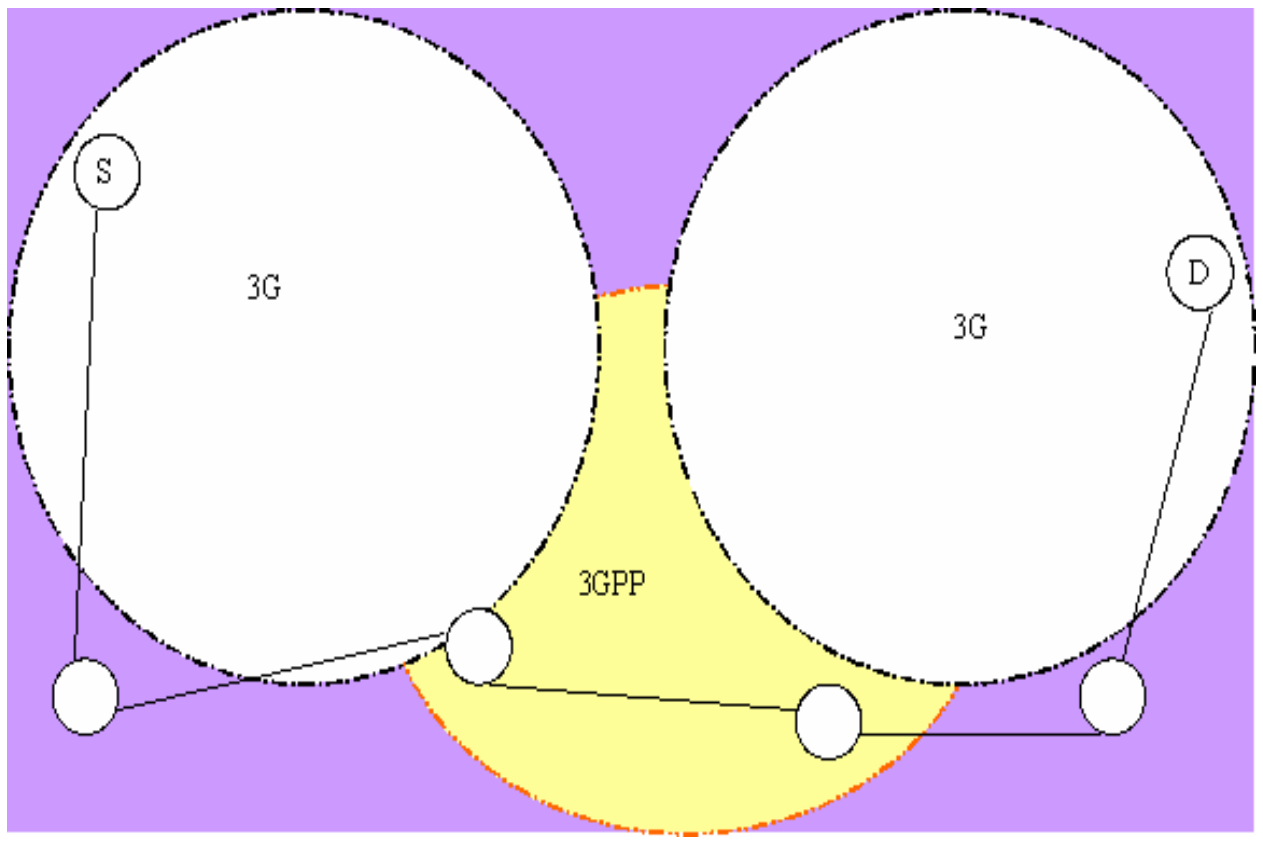

*Fig.4Path Representation between source (S) and Destination (D)*

- For all the network in the specified path between the source and destination
  - Get the bandwidth of the network
  - Calculate the average traffic of the network

- Note down the obstacles on the way
- Note down the active service providers with their coverage area
- Note down the type of network services [3G, 3GPP]

- Calculate the traffic density in the in the network
  NLc + ΔE ≤ NLth → take the same path
  NLc → Current Network Load
  ΔE → Estimated Increase in Load
  NLth → Network Threshold Load

- Generate a shortest path based on
  - The maximum signal coverage
  - Traffic density below the NLth
- Continue the transmission in the specified path
- Continue the same procedure till the source reaches the destination or the source is stable or not in mobility

**Assumptions made for the analysis**

Atleast one type of service provider is available within the limit

## V. PERFORMANCE EVALUATION

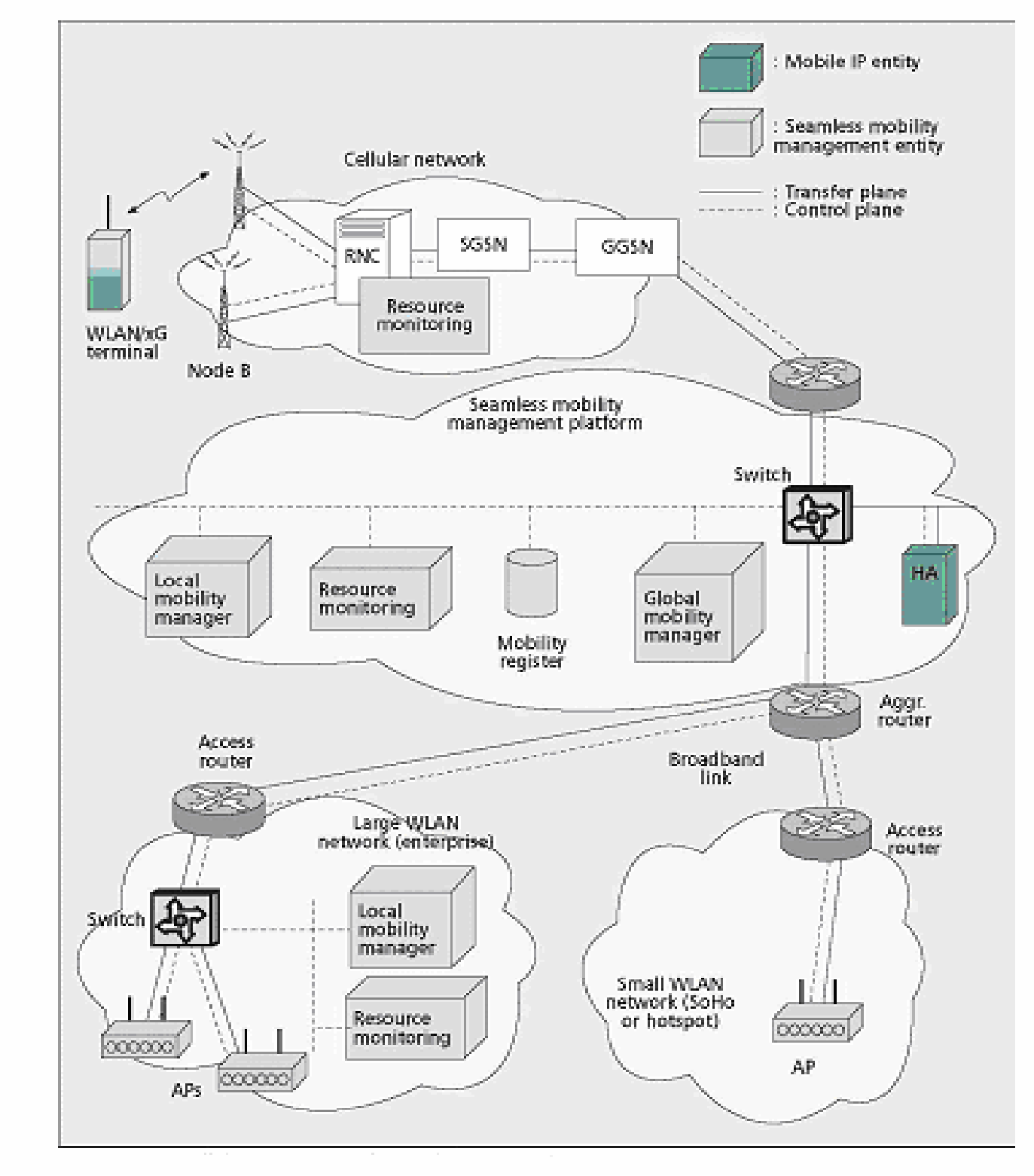

*Fig.5 Cellular/ WLAN Seamless Mobile Architecture[17]*

As far the architecture is considered the existing architecture is used without any modifications in the hardware. Modifications are carried out at the router level to reduce the router management delay as the information from the origin will have the path to be taken and the address of the destination also will have the details of the corresponding node through which it has to pass and when to go for handoff and what type of handoff is required in advance.

They may be triggered at application launch or during the application session by the terminal or local Mobility Management.

- The mobility register is a database that stores profiles and high-level characteristics of users, radio access networks, and operators. It also stores session information such as terminal location or application currently used. There is one database entity per operator.
- The global Mobility Management registers users to the seamless mobility service. The global Mobility Management activates the user context in the mobility register and is the screen for the Authentication Availability Access rights process. nterfaces with the mobility register receives "global" information from the terminal on application needs, user preferences, and basic access network data type, identity, availability and performs pre-selection of the access network. Using information from the terminal, the global Mobility Management provides the local Mobility Management with an ordered list of recommended access networks based on operator policies e.g., prioritize WLAN access for streaming applications, manage access privileges according to user's profile, user and network profiles, application quality of service (QoS) constraints (e.g., minimum bandwidth, latency guaranty), and basic radio information such as availability of access networks.
- The local Mobility Management receives measurements of the environment in its coverage area from access network RM entities access network load and quality information such as delay, packet loss ratio and from the terminal on an event-triggered or a periodic basis, processes application requests from the terminal in order to map QoS application needs to radio parameters (e.g., bandwidth vs. load on the different networks) within the service-to-radio mapping control function, Computes handoff triggers such as radio coverage and quality on the current network being below an acceptable threshold, current access network load being above a threshold, and modification of the network classification provided by the global Mobility Management. Also processes the global Mobility Management recommendation before selecting an access network for a user. This recommendation arises on a global Mobility Management event trigger or a terminal event trigger. Finally, makes the final handoff decision based on the radio triggers, as well as on the global Mobility Management recommendation, and orders the terminal to execute the handover.
- The terminal implements a seamless mobility application programming interface in charge of Computing handover triggers, related to coverage radio signal strength and signal quality in order to detect whether the radio bearer fulfills application needs in terms of link quality.
- Sending out radio signal strength and quality measurement for current and target access networks on an event-triggered based on local Mobility Management request or triggers below thresholds, but also on a periodic basis if the radio metrics remain below the threshold. If the signal is good enough for the terminal, no periodic measurement will be sent to the local Mobility Management unless the local Mobility Management sends out a specific request to get information detecting available networks in line of sight, Managing user preferences, Invoking services via service-to-radio mapping control, the API provides application needs and a preferred access network for this application. Then terminal pre-selection is confirmed or not by Mobility Managements according to access network condition. executing the handover upon local Mobility Management order depends on the link layer technology. There is no specific constraint on the mobility management protocol, which could be based on Internet Engineering Task Force specifications, for instance triggering the handoff in case it does not receive local Mobility Management orders in time or when only one target access network remains available. Optionally, managing operator policies and preferences so that it can efficiently make the handover decision by itself in the absence of a local Mobility Management recommendation.

The access network Resource Management entities. This entity is access-technology-specific as it interfaces each access network with the local Mobility Management Resource Managements receive real-time radio load-related information from the access network access point and use it to provide load indicators to the local Mobility Management in a *standardized* abstracted format.

## VI. RESULTS AND DISCUSSION

The following section is to provide the seamless flow of information in a practical context that addresses the integrating of cellular and WLAN access networks. In order to implement the different functions listed earlier, some initial technological choices need to be made. First, only intertechnology handoffs WLAN are considered in the seamless mobility architecture. Intratechnology handoffs are taken care of by technology-specific mechanisms. Then Mobile IP has been chosen as the L3 protocol for handoff execution in the proof of concept, and is used on top of either IPv4 or IPv6 in order to provide session continuity during intertechnology handoff. A clear separation of handoff decision and execution processes allows any evolution of IP protocols to minimize new care-of address configuration and rerouting latencies, for instance, to replace baseline Mobile IP without modifying the proposed architecture.

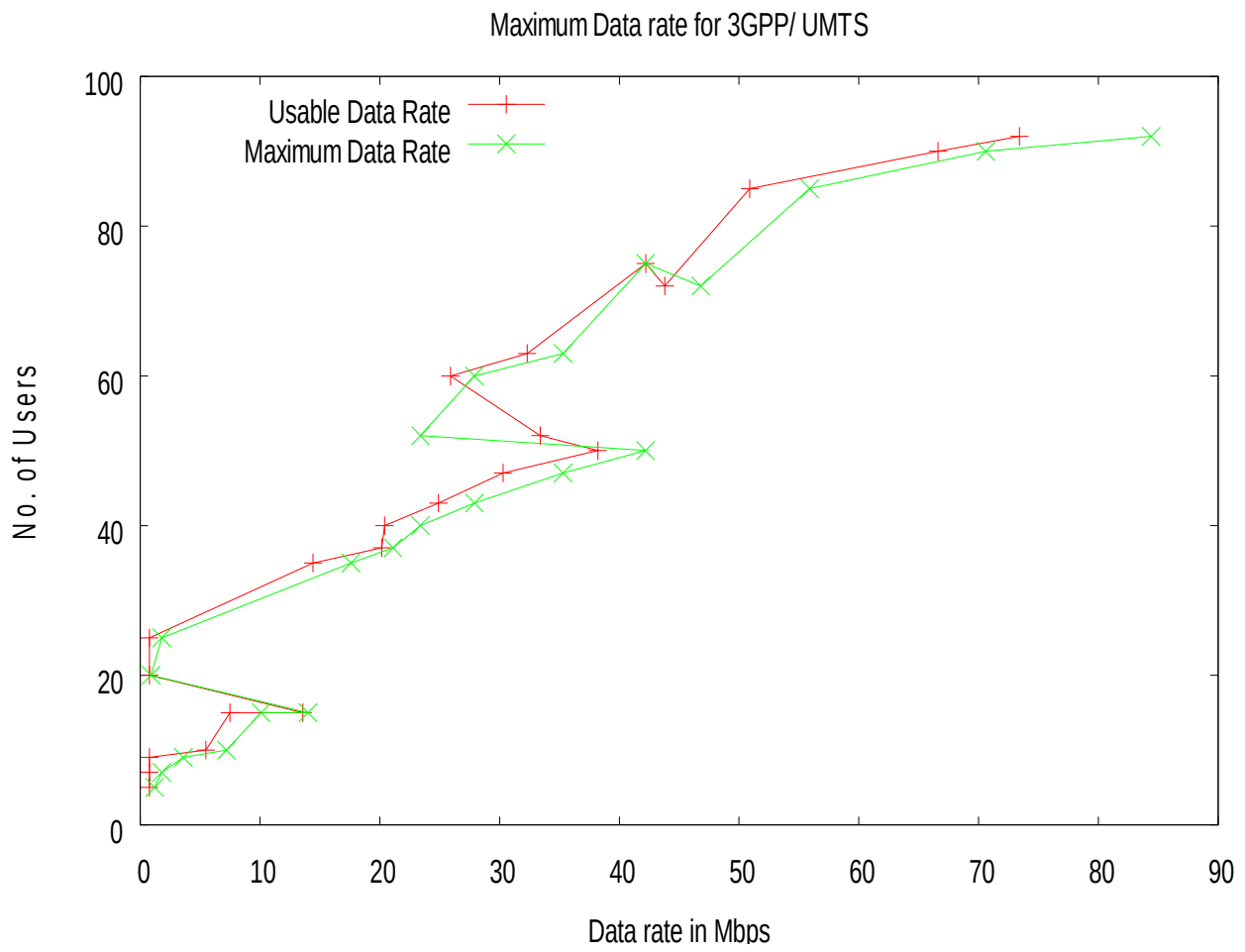

*Fig. 6. Data rate of the Network 3GPP/ UMTS*

AS the existing technology is considered the 3GPP/ UMTS network has a constant coverage for a limited traffic. The maximum users allowed is 80 – 85. Also as the usage spectrum is considered only 80 – 85% of the available spectrum is utilized efficiently. The result is shown in fig. 6. During which no traffic can be transmitted or received (it corresponds to traffic interruptions on Fig. 6. Upon returning to normal operation, a peak of traffic is observed when the terminal transmitting a burst of packets that could not be transmitted during the scanning periods of 200–400 ms. In order to avoid perceivable interruption of voice transmission, an adaptive buffer has been set up on the receiver side, which enables the network to cope with "silences" but results in slightly increased latency. This configuration could further be improved by breaking the scanning period into shorter ones in order to avoid latency increase. However, this configuration may lead to lower measurement precision, so an acceptable compromise must be reached. In any case, this scenario is not considered for wider-scale deployment, since the latency on the EDGE network leads to unacceptable VoIP quality.

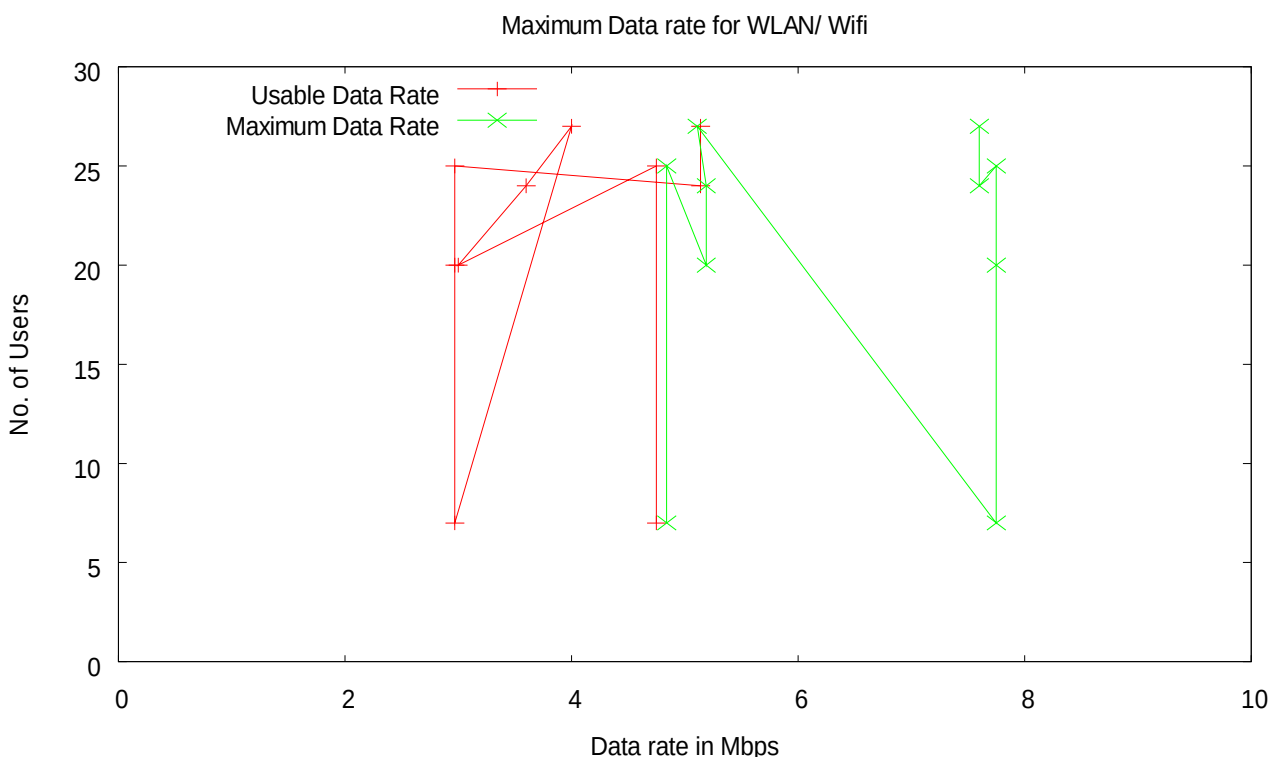

*Fig. 7. Data Rate of WLAN Network*

Another goal of the test bed was to assess performance of mobility management in the WLAN environment. As an example, we considered handoff delay for a 60 kb/s Real-Time Transmission Protocol streaming service, with handoff delay defined as the delay between the first Real-Time Transmission Protocol packet on the new interface and the last packet on the old interface. When network control is enabled, the decision to perform handover is taken on load criteria: the streaming starts on the WLAN interface where other terminals load the AP when the network load reaches a given threshold, mobility management entities trigger the handover. In both cases handoff delay was about 0.05 ms, because of Mobile IP and Generic Packet Radio Service network latencies. The results of the data rate in fig. 7 also gives a clear picture that it was mainly based on the nature of the application and also the stability of the network varies upon the nature of the functions of the hardware deployed.

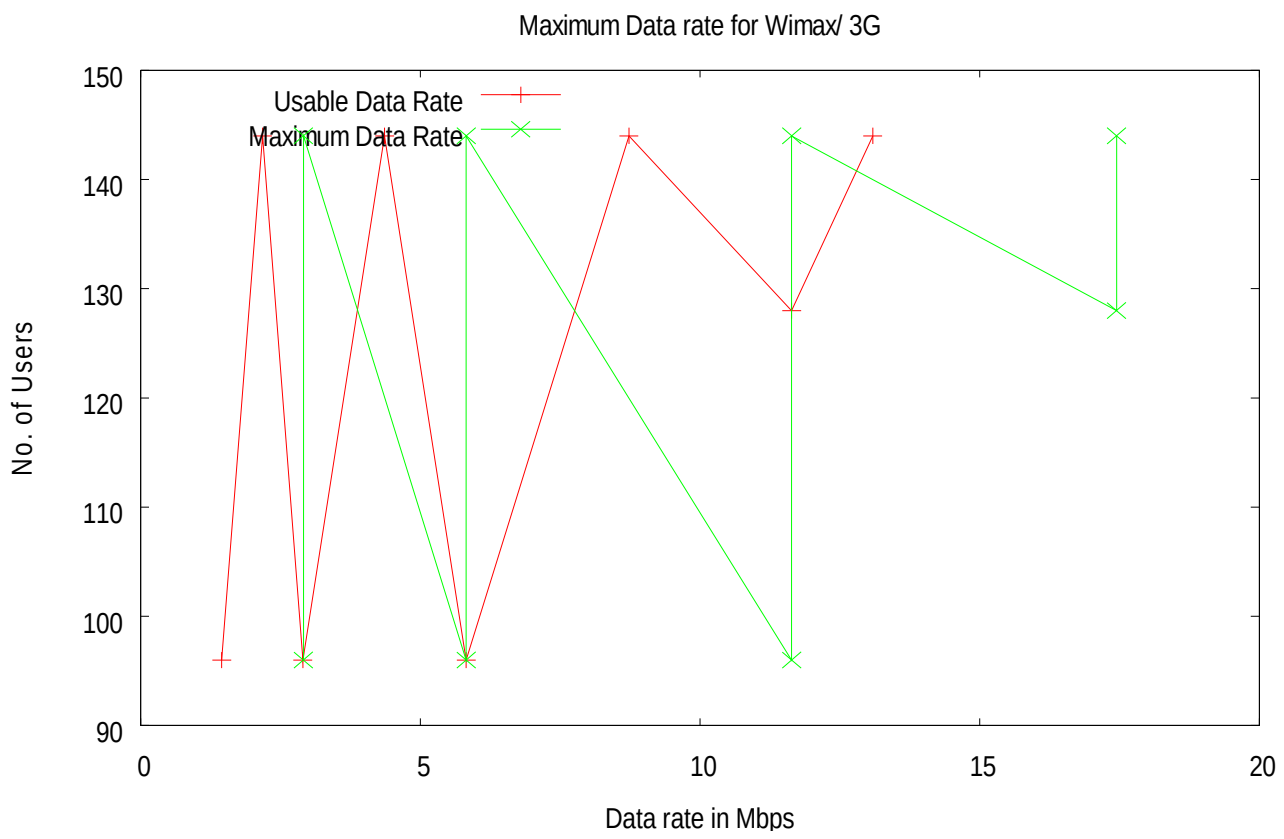

*Fig.8. Data Rate in the 3G Environment*

The higher transmission latency experienced in the cellular access network can be observed in the graph provided (fig. 8). On the transmit side, the transmission is performed with no silence period. On the receive side, handing over to the cellular network introduces more latency, results in a silence period the order of magnitude of which is equal to the latency difference between both networks. The use of an adaptive buffer at the receiver side makes it transparent to the user which is reflected as a smooth seamless flow in the heterogeneous Networks. When considering the 3G/ Wimax cellular network, the number of users is high compared with the other networks and also had a wider coverage but there is a pitfall at the end, the bandwidth fluctuates beyond 80%. At the time of mobility the network coverage is limited as shown in fig. 8

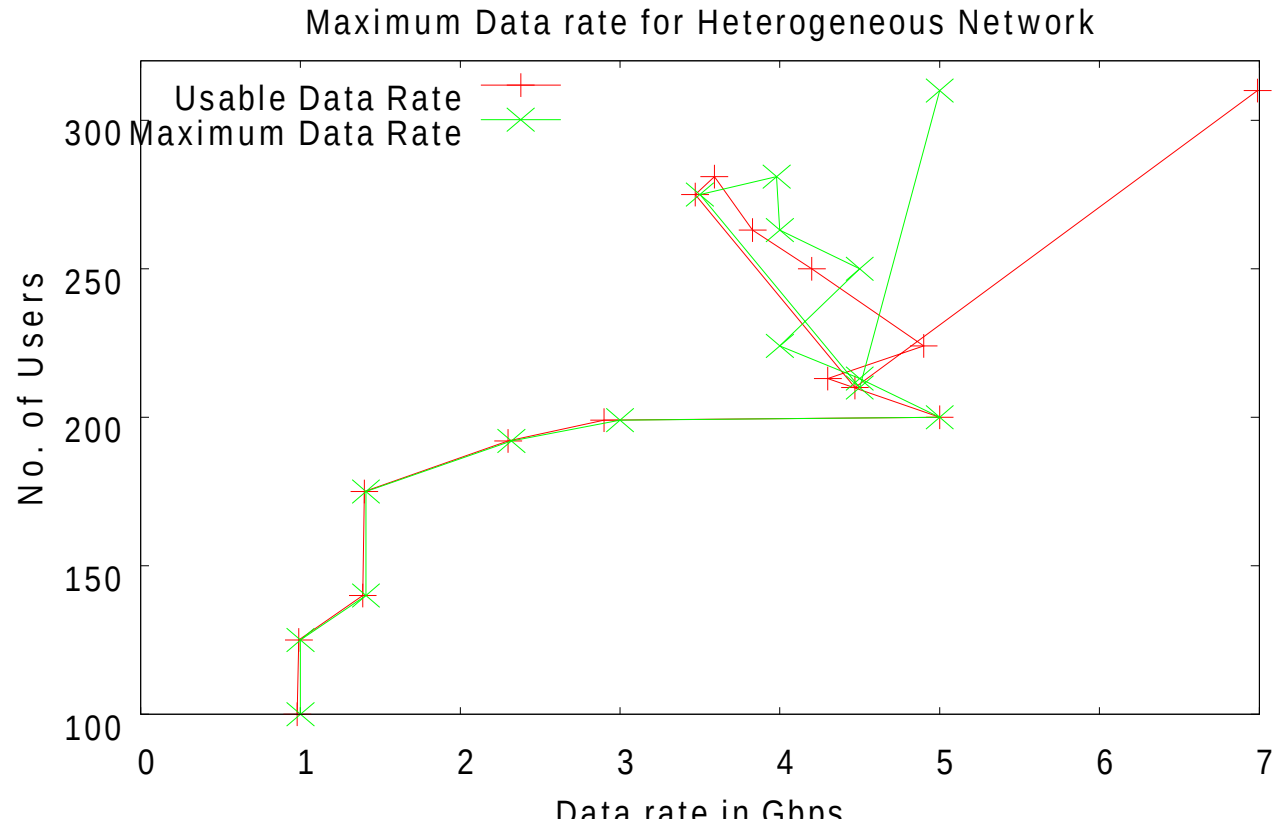

*Fig.9. Data Rate in the Heterogeneous Network Environment*

By considering the positive measures of the above mentioned networks and by having a thorough understanding between the available networks the heterogeneous network is designed. The heterogeneous network provides maximum throughput, minimum number of handoffs and maximum coverage at mobile. By designing a proper QoS standard and having proper understanding between the network the desires which are explained at the initial paragraph can be achieved. By improving the performance measures by deploying and allocating the code cpectrum for the 3GPP network and by having proper power management in the 3G network and by making use of antennas with wider coverage in WLAN environment, the available bandwidth can be maximum utilized and also the number of handoffs can be reduced as the nature of the network present in the graphical architecture between the source and the destination is studied in advance, a maximum throughput can be achieved with minimum tolerable delay or no delay based on the nature of the information that is taken for transmission. The data rate of the heterogeneous network is very close to the available rate as shown in fig. 9.

## VII. FUTURE SCOPE & CONCLUSION

Results have confirmed the feasibility of the approach; its scalability to large geographical areas has to be confirmed with additional validation through simulations and trials. A possible stepwise approach to the deployment of the different functional elements of the presented architecture is defined. In this approach a vector based location tracking and management is only considered for the seamless flow. By combining the parameters such as signal strength and delay management in flow and also the formats of the information we can have a seamless flow. Also the QoS between the Radio Access Networks's should be standardized in such a way that there is no mismatch of transmission from one type of environment to another type. Finally, with the advent of research on moving networks (e.g., Network Mobility), in which the whole network is mobile, the integration of WLANs and WMANs can improve mobile network connectivity. It is expected that public transportation (trains, buses, airplanes, ships, etc.) will popularize Internet access through WiFi connectivity to passengers while in movement. To this end, it will be equipped with a bridge to communicate with external networks such as WiMAX. Moreover, seamless mobility is still an issue as clients may be equipped with both interfaces, and the vehicle gateway may also give support to WiFi external connectivity through dual gateways /interfaces (WiFi/ WiMAX) in order to offer fault tolerance and load balance between networks as well as new connectivity opportunities to passengers. Apart from serving the movement network, the mobile gateway can also be used by external clients, such as those outside the WiFi AP and WiMAX BS coverage areas, but that have the opportunity to download data or attain Internet access through the dual gateway belonging to the vehicular area network (VAN).

## REFERENCES

[1] Fei Yu and Vikram Krishnamurthy, " Optimal Joint Session Admission Control in integrated WLAN and CDMA Cellular Networks with Vertical Handoff", IEEE Transaction on Mobile Computing, vol 6, No. 1, pp. 126 – 139, Jan' 2007.

[2] Jaroslav Holis and Pavel Pechac," *Elevation Dependent Shadowing Model for Mobile Communications via High Altitude Platforms in Built-Up Areas"* - IEEE TRANSACTIONS ON

ANTENNAS AND PROPAGATION, VOL. 56, NO. 4, APRIL 2008

[3] Sándor Imre, “Dynamic Call Admission Control for Uplink in 3G/4G CDMA-Based Systems” - IEEE TRANSACTIONS ON VEHICULAR TECHNOLOGY, VOL. 56, NO. 5, SEPTEMBER 2007

[4] Yan Zhang, and Masayuki Fujise, “Location Managemenet Congestion Problem in Wireless Networks” - IEEE TRANSACTIONS ON VEHICULAR TECHNOLOGY, VOL. 56, NO. 2, MARCH 2007

[5] Daniel Morris and A. Hamid Aghvami, “Location Management Strategies for Cellular Overlay Networks—A Signaling Cost Analysis” - IEEE TRANSACTIONS ON BROADCASTING, VOL. 53, NO. 2, JUNE 2007

[6] Haining Chen, Hongyi Wu, Sundara Kumar and Nian-Feng Tzeng, “Minimum-Cost Data Delivery in Heterogeneous Wireless Networks” - IEEE TRANSACTIONS ON VEHICULAR TECHNOLOGY, VOL. 56, NO. 6, NOVEMBER 2007

[7] Yuh-Shyan Chen, Ming-Chin Chuang, and Chung-Kai Chen, “DeuceScan: Deuce-Based Fast Handoff Scheme in IEEE 802.11 Wireless Networks” - IEEE TRANSACTIONS ON VEHICULAR TECHNOLOGY, VOL. 57, NO. 2, MARCH 2008

[8] Mohammed M. Olama, Seddik M. Djouadi, Ioannis G. Papageorgiou, and Charalambos D. Charalambous “Position and Velocity Tracking in Mobile Networks Using Particle and Kalman Filtering With Comparison” - IEEE TRANSACTIONS ON VEHICULAR TECHNOLOGY, VOL. 57, NO. 2, MARCH 2008

[9] Archan Misra, Abhishek Roy and Sajal K. Das, “Information-Theory Based Optimal Location Management Schemes for Integrated Multi-System Wireless Networks” - IEEE/ACM TRANSACTIONS ON NETWORKING, VOL. 16, NO. 3, JUNE 2008

[10] Yang Xiao, Yi Pan and Jie Li, “Design and Analysis of Location Management for 3G Cellular Networks” - IEEE TRANSACTIONS ON PARALLEL AND DISTRIBUTED SYSTEMS, VOL. 15, NO. 4, APRIL 2004

[11] Di-Wei Huang, Phone Lin and Chai-Hien Gan, “Design and Performance Study for a Mobility Management Mechanism (WMM) Using Location Cache for Wireless Mesh Networks” -IEEE TRANSACTIONS ON MOBILE COMPUTING, VOL. 7, NO. 5, MAY 2008

[12] Yi-hua Zhu and Victor C. M. Leung, “Optimization of Sequential Paging in Movement-Based Location Management Based on Movement Statistics” - IEEE TRANSACTIONS ON VEHICULAR TECHNOLOGY, VOL. 56, NO. 2, MARCH 2007

[13] Ramón M. Rodríguez-Dagnino, and Hideaki Takagi, “Movement-Based Location Management for General Cell Residence Times in Wireless Networks” - IEEE TRANSACTIONS ON VEHICULAR TECHNOLOGY, VOL. 56, NO. 5, SEPTEMBER 2007

[14] Wenchao Ma, Yuguang Fang and Phone Lin, “Mobility Management Strategy Based on User Mobility Patterns in Wireless Networks” - IEEE TRANSACTIONS ON VEHICULAR TECHNOLOGY, VOL. 56, NO. 1, JANUARY 2007

[15] Dusit Niyato and Ekram Hossain, “A Non-cooperative Game-Theoretic Framework for Radio Resource Management in 4G Heterogeneous Wireless Access Networks” - IEEE TRANSACTIONS ON MOBILE COMPUTING, VOL. 7, NO. 3, MARCH 2008

[16] Marcus L. Roberts, Michael A. Temple, Richard A. Raines, Robert F. Mills, and Mark E. Oxley, “Communication Waveform Design Using an Adaptive Spectrally Modulated, Spectrally Encoded (SMSE) Framework” - IEEE JOURNAL OF SELECTED TOPICS IN SIGNAL PROCESSING, VOL. 1, NO. 1, JUNE 2007

[17] Christian Makaya and Samuel Pierre, “An Architecture for Seamless Mobility Support in IP-Based Next-Generation Wireless Networks” - IEEE TRANSACTIONS ON VEHICULAR TECHNOLOGY, VOL. 57, NO. 2, MARCH 2008

[18] Haiyun Luo, Xiaqiao Meng, Ram Ramjee, Prasun Sinha, and Li (Erran) Li, “The Design and Evaluation of Unified Cellular and Ad Hoc Networks” - IEEE TRANSACTIONS ON MOBILE COMPUTING, VOL. 6, NO. 9, SEPTEMBER 2007\

[19] Anthony Almudevar, “Approximate Calibration-Free Trajectory Reconstruction in a Wireless Network” - IEEE TRANSACTIONS ON SIGNAL PROCESSING, VOL. 56, NO. 7, JULY 2008

[20] Yi Yuan-Wu and Ye Li, “Iterative and Diversity Techniques for Uplink MC-CDMA Mobile Systems With Full Load” - IEEE TRANSACTIONS ON VEHICULAR TECHNOLOGY, VOL. 57, NO. 2, MARCH 2008

[21] Abhishek Roy, Archan Misra, and Sajal K. Das, “Location Update versus Paging Trade-Off in Cellular Networks: An Approach Based on Vector Quantization” - IEEE TRANSACTIONS ON MOBILE COMPUTING, VOL. 6, NO. 12, DECEMBER 2007

[22] Enrique Stevens-Navarro, Yuxia Lin, and Vincent W. S. Wong, “An MDP-Based Vertical Handoff Decision Algorithm for Heterogeneous Wireless Networks” - IEEE TRANSACTIONS ON VEHICULAR TECHNOLOGY, VOL. 57, NO. 2, MARCH 2008

[23] Sai Shankar N and Mihaela van der Schaar, “Performance Analysis of Video Transmission Over IEEE 802.11a/e WLANs”- IEEE TRANSACTIONS ON VEHICULAR TECHNOLOGY, VOL. 56, NO. 4, JULY 2007

[24] Abhishek Roy, Archan Misra and Sajal K. Das, “Location Update versus Paging Trade-Off in Cellular Networks: An Approach Based on Vector Quantization” - IEEE TRANSACTIONS ON MOBILE COMPUTING, VOL. 6, NO. 12, DECEMBER 2007

[25] Dusit Niyato, and Ekram Hossain, “A Noncooperative Game-Theoretic Framework for Radio Resource Management in 4GHeterogeneous Wireless Access Networks” - IEEE TRANSACTIONS ON MOBILE COMPUTING, VOL. 7, NO. 3, MARCH 2008